\documentclass
[aps,prd,nofootinbib,
superscriptaddress,preprintnumbers,
twocolumn
]{revtex4-1}

\usepackage{hyperref}
\usepackage{color}
\usepackage[dvipsnames]{xcolor}%
\usepackage{graphicx,subfigure}
\usepackage{siunitx}
\usepackage{amsmath,amssymb,amsfonts}
\usepackage{bm}
\usepackage{hyperref}
\usepackage{lineno}
\usepackage{soul}
\usepackage{enumitem}
\usepackage{changepage}

\graphicspath{{./plot/}}

\def \be {\begin{equation}}
\def \ee {\end{equation}}
\def \bes {\begin{subequations}}
\def \ees {\end{subequations}}

\newcommand{\beq}{\begin{eqnarray}}
\newcommand{\eeq}{\end{eqnarray}}

\def \pd {\partial}
\newcommand{\Eq}[1]{Eq.~\eqref{#1}}
\newcommand{\Fig}[1]{Fig.~\ref{#1}}

\newcommand{\sH}{\mathcal{H}}

\def \le<{\langle}
\def \ri>{\rangle}
\def \+{\dagger}
\def \le({\left(}
\def \ri){\right)}
\def \le[{\left[}
\def \ri]{\right]}

\def \e {\epsilon}

\def \vp {\bm{p}}

\def \hp {\hat{p}}

\def \vz {\hat{p}_z}

\newcommand\sect[1]{\noindent \textbf{#1.}---}

\begin{document}

\preprint{CERN-TH-2022-196, LA-UR-22-32315}

\title{
%
Far-from-equilibrium slow modes and momentum anisotropy in expanding plasma
}

\author{Jasmine Brewer}
\affiliation{Theoretical Physics Department, CERN, CH-1211 Gen\`eve 23, Switzerland}

\author{Weiyao Ke}
\affiliation{Theoretical Division, Los Alamos National Laboratory, Los Alamos NM 87545, United States}

\author{Li Yan}
\affiliation{Institute of Modern Physics, Fudan University, Handan Road 220, Yangpu District, Shanghai,
200433, China}

\author{Yi Yin
\footnote{yiyin@impcas.ac.cn}
}
\affiliation{Quark Matter Research Center, Institute of Modern Physics, Chinese Academy of Sciences, Lanzhou 730000, China}

\date{\today}

\begin{abstract}

The momentum distribution of particle production in heavy-ion collisions encodes information about thermalization processes in the early-stage quark-gluon plasma.
We use kinetic theory to study the far-from-equilibrium evolution of an expanding plasma with an anisotropic momentum-space distribution.
We identify slow and fast degrees of freedom in the far-from-equilibrium plasma from the evolution of moments of this distribution.
At late times, the slow modes correspond to hydrodynamic degrees of freedom and are naturally gapped from the fast modes by the inverse of the relaxation time, $\tau_R^{-1}$. At early times, however, there are an infinite number of slow modes
with a gap inversely proportional to time, $\tau^{-1}$. 
From the evolution of the slow modes we generalize the paradigm of the far-from-equilibrium attractor to vector and tensor components of the energy-momentum tensor, and even to higher moments of the distribution function that are not part of the hydrodynamic evolution.
We predict that initial-state momentum anisotropy decays slowly in the far-from-equilibrium phase and may persist until the relaxation time.

\end{abstract}

\maketitle

The rich dynamics of many-body quantum systems out of equilibrium poses many interesting outstanding challenges and is an active area of research in cold atomic gases and the quark--gluon plasma produced in heavy-ion collisions.
Though far-from-equilibrium systems are generally extremely complicated, with degrees of freedom commensurate with the number of microscopic states in the system, in some cases the dynamics can be described by a reduced number of effective degrees of freedom. 
Hydrodynamics is an example of such an effective theory for many-body systems near equilibrium,
in which the relevant (slow) degrees of freedom are associated to conserved densities.
Out of equilibrium, the dynamics of a system may also simplify substantially near non-thermal fixed points~\cite{Berges:2008wm}.
Identifying possible slow degrees of freedom (modes) remains a major challenge in the study of far-from-equilibrium systems in general~\cite{non-eq-review, Brauner:2022rvf}, and the quark--gluon plasma (QGP) in particular~\cite{Berges:2020fwq,Romatschke:2017ejr,Spalinski:2022cgj}.

Though hydrodynamic modelling has been enormously successful in heavy-ion collisions~\cite{Shen:2020mgh}, the initial stages of the collision are very far from equilibrium and therefore beyond the expected regime of applicability of hydrodynamics. Smaller collision systems (e.g. proton--ion and light-ion) may be out of equilibrium for their whole lifetime~\cite{Brewer:2021kiv}.
Remarkably, in many scenarios hydrodynamics describes the evolution of some macroscopic quantities even far-from-equilibrium~\cite{Chesler:2009cy,Heller:2011ju,Chesler:2015wra,Chesler:2015bba,Casalderrey-Solana:2013sxa,Casalderrey-Solana:2013aba,Keegan:2015avk} and can be formulated for highly anisotropic systems~\cite{Martinez:2010sc,Florkowski:2010cf,Alqahtani:2017mhy}.
Importantly, in a class of models of different microscopic theories, the far-from-equilibrium evolution of the plasma rapidly loses memory 
of different initial conditions and macroscopic quantities exhibit universal 
attractive behavior~\cite{Heller:2015dha,Heller:2016rtz,Florkowski:2017olj,Romatschke:2017vte,Spalinski:2017mel, Romatschke:2017acs,Behtash:2017wqg,Strickland:2017kux,Behtash:2018moe,Strickland:2018ayk,Heller:2018qvh,Blaizot:2019scw, Kurkela:2019set,Strickland:2019hff,Heller:2020anv,Almaalol:2020rnu,Alalawi:2022pmg}. 
These findings suggest that 
the far-from-equilibrium QGP can be characterized through the reduced degrees of freedom of the attractor, with deviations from the attractor decaying quickly. 
Macroscopic quantities far-from-equilibrium follow the attractor 
towards the hydrodynamic solution at late times, which suggests a connection between the hydrodynamic degrees of freedom and the attractor.

However, studies so far have focused on the scalar components 
of the energy-momentum tensor, such as energy density $\e$ and longitudinal and transverse pressures $p_L$ and $p_T$.
The behavior of the other components
has rarely been investigated. Consequently, a general relation between far-from-equilibrium slow modes and hydrodynamic degrees of freedom remains elusive.
In this letter, we tackle this important gap and reveal important theoretical and phenomenological consequences of the evolution of these components.

For a longitudinally expanding plasma that is spatially homogeneous but has general shape in momentum space, we identify the underlying slow modes using the adiabatic analysis of kinetic theory developed in~\cite{Brewer:2019oha,Blaizot:2019scw,Brewer:2022vkq}. 
Unlike for the scalar components of the energy-momentum tensor, we find that the 
early-time slow modes are not related to hydrodynamic modes in general, until the relaxation time. 
At early times the fast longitudinal expansion drives a tower of slow degrees of freedom that do not evolve into hydrodynamic modes.
This implies that hydrodynamization, namely the dominance of hydrodynamic degrees of freedom, does not occur far from equilibrium. 

Key to understanding equilibration processes in the quark--gluon plasma from phenomenology is understanding whether the
anisotropic flow observed in small collision systems is hydrodynamic in origin or receives
significant contributions from the momentum anisotropy generated by quantum fluctuations in the initial state~\cite{Dumitru:2008wn,Dusling:2015gta,Schlichting:2016sqo,Schenke:2019pmk}. In far-from-equilibrium systems the time scale for the decay of initial-state momentum anisotropies is under debate~\cite{Greif:2017bnr,Nie:2019swk,Liyanage:2022nua}.
Based on the properties of early-time slow modes, 
we find that the initial momentum anisotropy decays more slowly far from equilibrium than would be expected based on extending hydrodynamics, and could persist throughout the lifetime of small collision systems.

\sect{Kinetic description}
We consider the angular 
distribution $F(\tau, \hp)=\frac{1}{2\pi^2}\int dp\, p^{3}\, f(\tau, \vp)$ of the (massless) gluon distribution $f$,
which characterizes the 
evolution the energy-momentum tensor 
and moments of the same dimension (see Eq.~\ref{moments} below).
For a longitudinally-expanding system 
with spatial homogeneity,
$F(\tau, \hat p)$
satisfies
~\cite{Kurkela:2019kip}
\begin{align}
\label{F-evo}
\pd_{\tau} F 
- \frac{\vz(1-\vz^2)}{\tau}\pd_{\vz}\,F +\frac{4\vz^{2}}{\tau}F
= - C[f]\, 
\end{align}
with $\hat{p}_z \equiv p_z/p$ and $\tau=\sqrt{t^2-z^2}$ the proper time. 
With Eq.~\eqref{F-evo}, we focus our study on the evolution of anisotropies in the momentum distribution
and will estimate the role of spatial gradients separately (see discussion later, or as in Ref.~\cite{Kurkela:2019kip}).
For illustrative purposes we will close \Eq{F-evo} by assuming the collision integral is only a function of $F$, and takes the form~\cite{Abrikosov_1959,Rocha:2021zcw}
\begin{equation}
\label{RTA}
C[F]
=-\frac{F-T^{00}-3(T^{0x}\hat{p}_x+ T^{0y}\hat{p}_{y}+\tau T^{0\eta}\hat{p}_{z})}{\tau_{R}}\,, 
\end{equation}
where $F$ relaxes to its equilibrium form
on the timescale $\tau_R$.
Eq.~\ref{RTA} conserves energy and momentum density.

\sect{Moment equations}
We first expand $F$ in spherical harmonics and recast Eq.~\eqref{F-evo} into evolution equations for the moments
\begin{align}
\label{moments}
    L_{lm}&\equiv
    \int^{1}_{-1} 
    d\vz
    \int
    \frac{d\phi}{4\pi} 
    F(\tau,\hat{p})(-1)^{m} P^{m}_{l}(\vz)\cos(m\phi)\,. 
\end{align}
Here $P^{m}_{l}$ are the standard associated Legendre polynomials and $\tan\phi \equiv \hat p_y/\hat p_x$.
Moments obtained from \Eq{moments} by replacing $\cos(m\phi) \rightarrow \sin(m\phi)$ are equivalent due to the rotational symmetry in the transverse plane so we do not consider them separately.
By construction, all independent components in $T^{\mu\nu}$ are related to the moments with $l\leq 2$, i.e., $L_{00} = T^{00} = \epsilon$, $L_{10} = \tau T^{0\eta}$, $L_{11} = T^{0x}$, $L_{20} = \frac{1}{2}(3\e-p_{L})$ and $L_{22} = 3 \left(T^{xx}-T^{yy}\right)$.

\Eq{F-evo} with the collision integral \Eq{RTA} 
is invariant under $SO(2)$ rotations in the transverse plane and a $Z_2$ ``parity" transformation $p_{z}\to -p_{z}$.
Since $L_{lm}\to (-1)^{l+m}L_{lm}$ under the parity transformation, we use $s=+$($-$) to denote moments with even(odd) $l+m$.
Moments with different $(m,s)$ decouple from one another and evolve independently, so we organize all moments into a vector for each $(m,s)$, $\psi_{ms}=(L_{m+\delta_{s-},m}, L_{m+\delta_{s-}+2,m},\ldots)$.
Substituting \Eq{moments} into Eq.~\eqref{F-evo} yields an evolution equation for moments in each $(m,s)$ sector
\begin{equation}
\label{psims}
\pd_{\tau}\psi_{ms} =- H_{ms}(\tau) \,\psi_{ms}\,.
\end{equation}
The matrix $H_{ms}$ is obtained from the properties of spherical harmonics and is given in the Supplemental Material.

\begin{figure*}
	\centering
        \includegraphics[width=\linewidth]{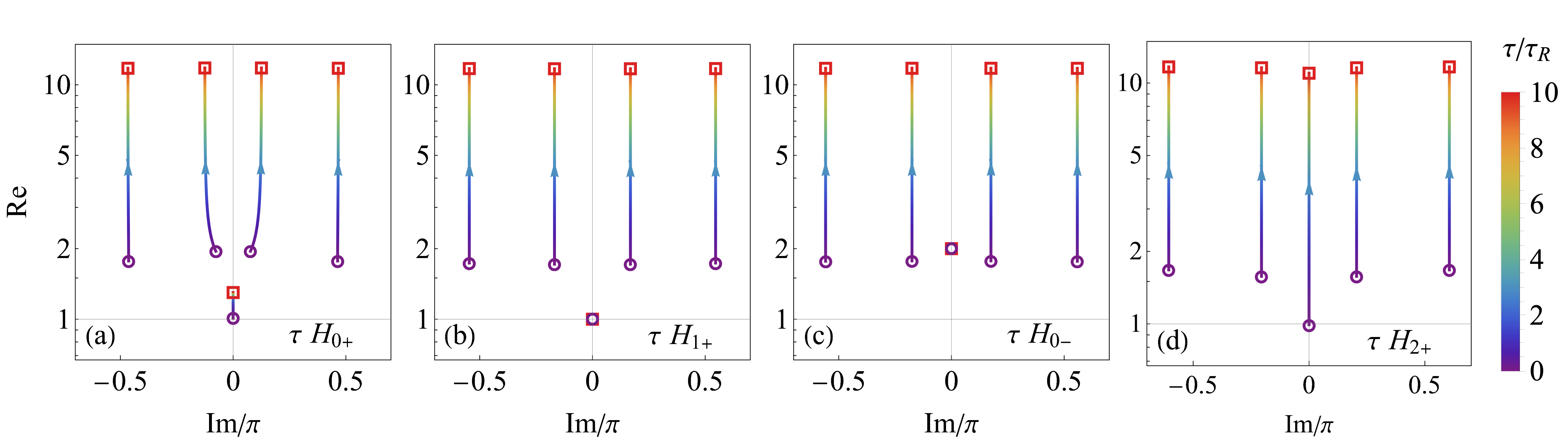}
 	\caption{
 	\label{fig:eigenvalues}
The time evolution of five low-lying eigenvalues of $\tau H_{ms}$ in the complex plane for illustrative sectors $(m,s)=(0,+)$ (a),$(1,+)$ (b), $(0,-)$ (c) and $(2,+)$ (d). Purple circles and red squares show the eigenvalues of $\tau H_{ms}$ at $\tau=0$ and $\tau=10\tau_R$, respectively.
	}
\end{figure*}

\begin{figure*}[t!]
	\centering
        \includegraphics[width=0.245\linewidth]{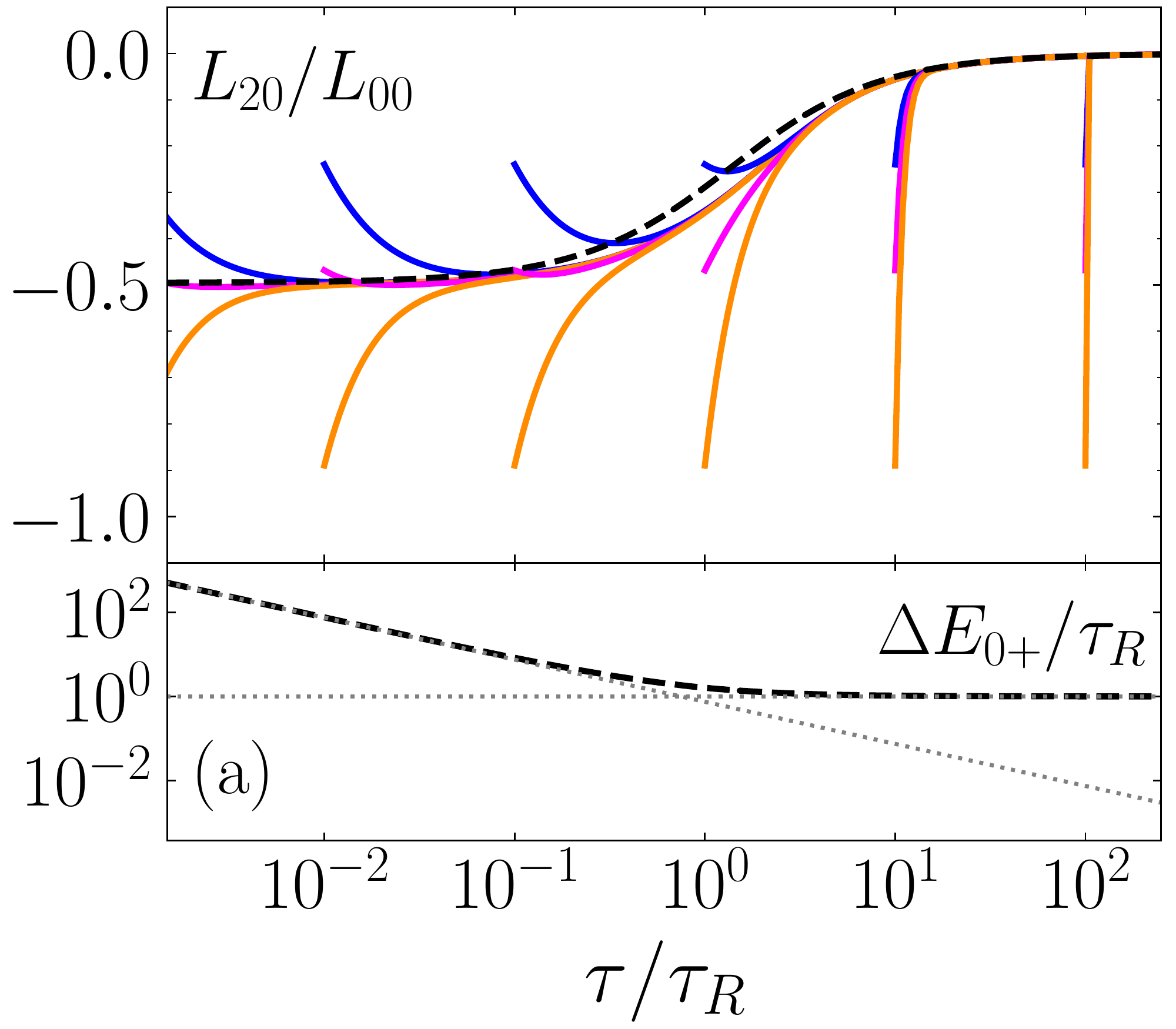}
        \includegraphics[width=0.245\linewidth]{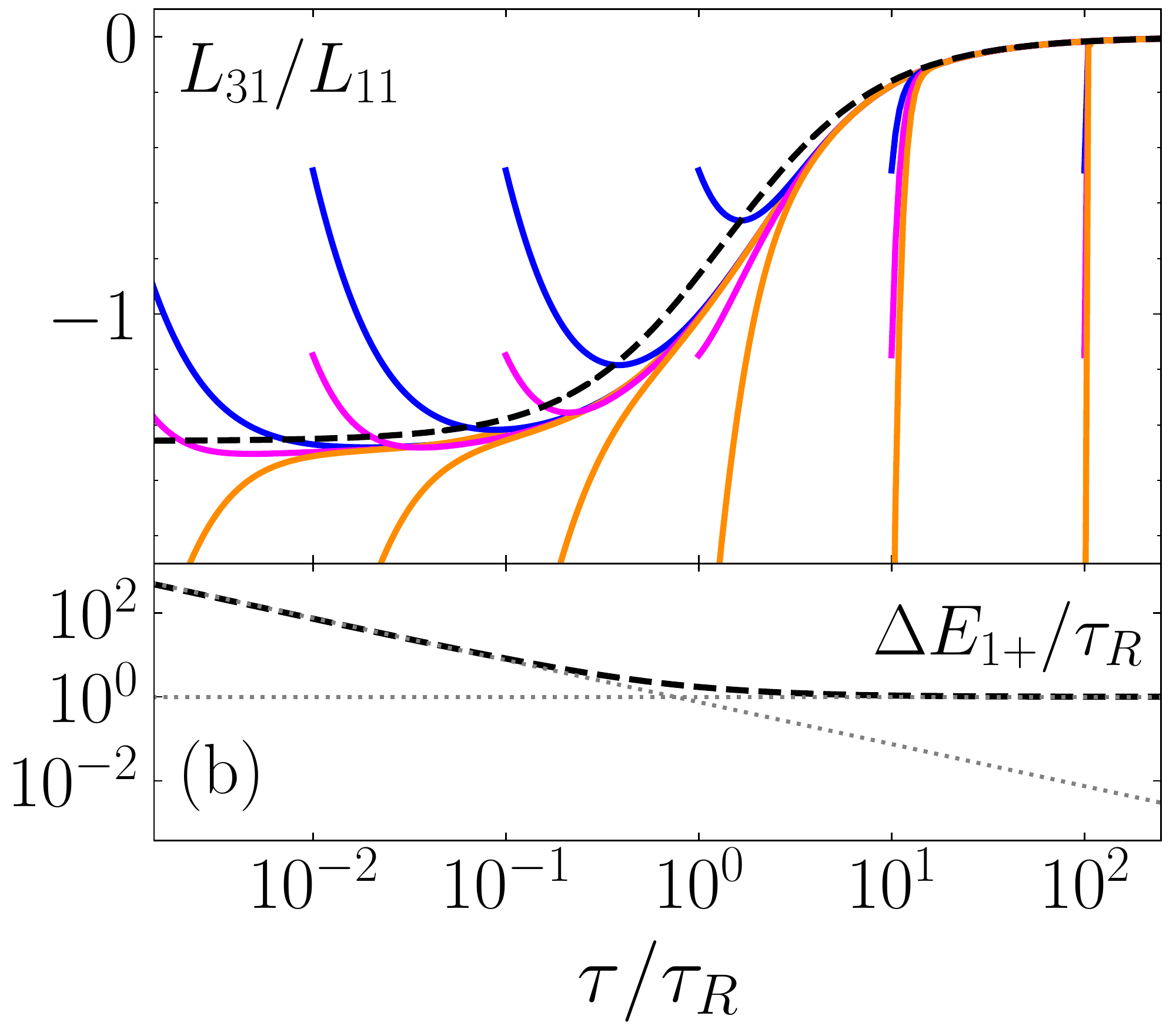}
        \includegraphics[width=0.245\linewidth]{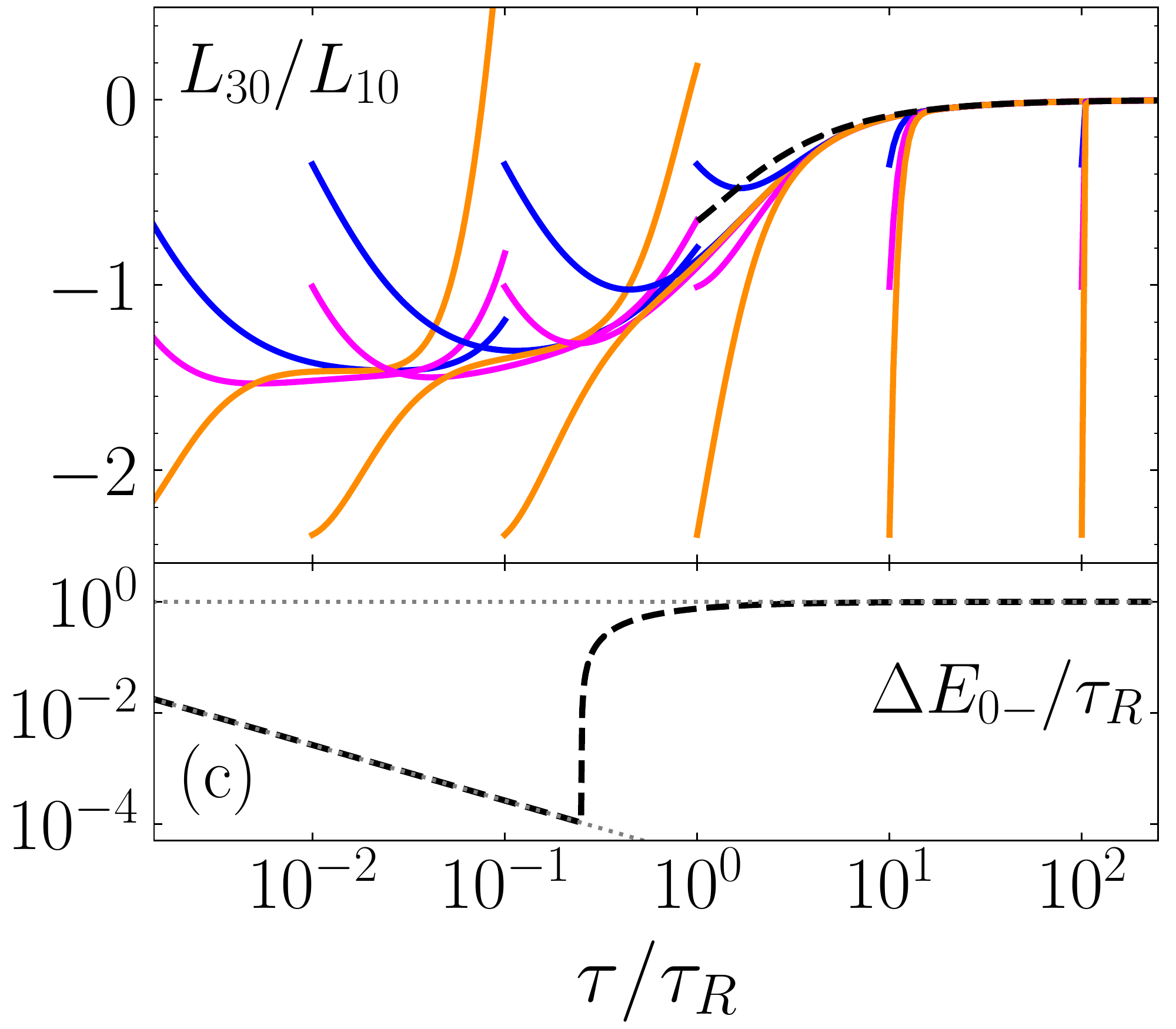}
\includegraphics[width=0.245\linewidth]{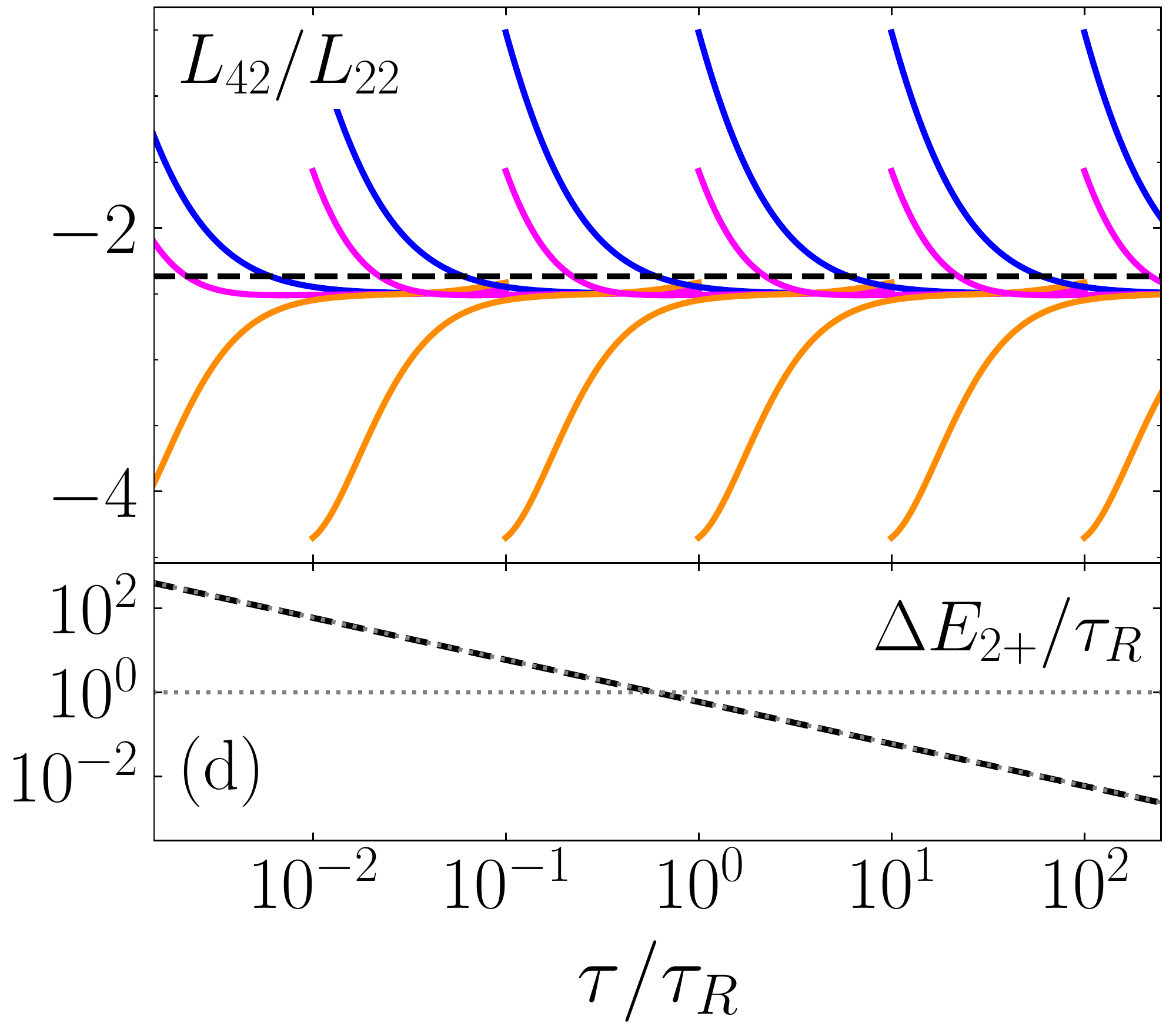}
 	\caption{
 	\label{fig:attractor}
Upper panels: The time evolution of the ratio of the two lowest moments in each of the sectors $(m,s)=(0,+)$ (a),$(1,+)$ (b), $(0,-)$ (c) and $(2,+)$ (d) from \Fig{fig:eigenvalues}. Colors show different initial conditions for the distribution function from varying $\alpha_n$, and the dashed black curve is the ratio of moments for the ground state of $H_{ms}$. Lower panels: The evolution of the energy gap of $H_{ms}$ in dashed black, compared to references of $1/\tau$ and $1$ in grey. The discontinuity of the gap in (c) is from the crossing of the real and complex eigenvalues (see \Fig{fig:eigenvalues}c).
	}
\end{figure*}

\sect{Slow modes}
To identify slow modes in the out-of-equilibrium evolution, we analyze the eigenspectrum of the non-Hermitian matrices $H_{ms}(\tau)$. If $H_{ms}$ is truncated as an $N \times N$ matrix, there are $N$ eigenstates for each $(m,s)$, $\phi^N_{ms}$, which we order by the real part of their eigenvalue, $E^N_{ms}$, since it sets their decay rate.
In each sector we denote the slowest eigenmode as the ground state $\phi^{G}_{ms}$.

For illustrative purpose we will focus on sectors containing relevant components of $T^{\mu\nu}$. Sectors $(m,s)=(0,+), (1,+)$, and $(0,-)$ contain moments corresponding to the 
energy density $\e$ and momentum densities $P^{x}=T^{0x}$ and $P^{z}=\tau T^{0\eta}$, respectively, and therefore will be referred to as \textit{hydrodynamic sectors}. 
We will additionally consider the $(2,+)$ sector since it contains the tensor part $T^{ij}$ and hence information about the momentum anisotropy.
In Fig.~\ref{fig:eigenvalues} we show the evolution of the lowest five eigenvalues in each sector 
as a function of time from $\tau=0$ (purple circles) to $\tau/\tau_R=10$ (red squares), obtained from truncating $H_{ms}$ at $N=11$.

Though most eigenvalues are complex with approximately degenerate real parts, 
each sector has at least one 
real eigenmode. 
In the {\it hydrodynamic sectors} we denote the real modes by $\phi^{H}$. These modes are generalized hydrodynamic modes
since they evolve continuously into the hydrodynamic degrees of freedom at late times, namely,
$\phi^{H}_{0+}\to(\e,0,\ldots)$, $\phi^{H}_{1+}\to(T^{0x},0,\ldots)$ and $\phi^H_{0-}\to (\tau T^{0\eta},0,\ldots)$. 
At late times the $\phi^H$ are the slowest modes because 
their eigenvalues 
are gapped from eigenvalues of other modes
with a gap $1/\tau_R$ (see red squares in Fig.~\ref{fig:eigenvalues}(a)-(c)). 
The eigenvalues of $\phi^H$ are given by the conservation equation for $T^{\mu\nu}$, e.g., 
$\pd_{\tau}P^{x}/P^{x}=-1/\tau,
\pd_{\tau}P^{z}/P^{z}=-2/\tau
$. 
The hydrodynamic modes survive in the long time limit while the gapped non-hydrodynamic modes vanish on a time scale $\tau_R$ (cf. Ref~\cite{Romatschke:2017ejr}).

However, hydrodynamic modes do {\it not} in general emerge smoothly from the slow modes from far from equilibrium.
This can be seen from the evolution of the generalized hydrodynamic modes for $\tau \ll \tau_R$ 
(purple circles in Fig.~\ref{fig:eigenvalues}). 
While 
the generalized hydrodynamic modes in the parity-even sectors $\phi^{H}_{0+}$ and $\phi^{H}_{1+}$ remain gapped from excited states, 
the parity-odd mode $\phi^{H}_{0-}$ is not the 
slowest eigenmode until the level crossing in the $(0,-)$ sector occurs around $\tau\sim \tau_R$. When $\tau<\tau_R$
the ground state in the $(0,-)$ sector is complex and degenerate. This demonstrates that hydrodynamic modes are not necessarily the slow modes out of equilibrium.

In the non-hydrodynamic sector $(2,+)$ the real mode is the ground state $\phi^{G}_{2+}$ 
and it evolves in time. Once slow modes dominate, their features describe the evolution of the shape of the phase space distribution.
In the Supplemental Material we show that the ground state eigenvalues for all $m \geq 2$ in the parity-even sector are
\begin{equation}
\label{Em}
     E^{G}_{m+}=\frac{1}{\tau}+\frac{1}{\tau_{R}}\, ,\quad
    m\geq 2\, .
\end{equation}
Remarkably, 
the ground state eigenvalues are identical for all parity-even sectors at early times,
\begin{align}
\label{E-slow}
    \lim_{\tau\to 0}\, E^{H}_{0+}=E^{H}_{1+}=E^{G}_{m\geq2 +}=\frac{1}{\tau} \,\,,
\end{align}
which is a consequence of the free-streaming expansion.

\sect{Attractor}
Once faster modes have decayed, the evolution of a system is captured by the evolution of the remaining slow modes.
For our purpose this means that the moments in each $(m,s)$ sector are fixed by the ground state $\phi^G_{ms}$ and do not depend on the initial distribution.
The dominance of slow modes thus leads to attractive behavior, as first discussed in Refs.~\cite{Brewer:2019oha,Brewer:2022vkq}. 
We now examine the dominance of slow modes by solving 
\Eq{psims} numerically.
We choose an initial distribution function inspired by Color Glass Condensate calculations, but with additional momentum anisotropy introduced through coefficients $\alpha_n$:
\begin{align}
\label{eq:fI}
f_{I}(\vp)&= \frac{Q_{0}}{p} \frac{e^{-\frac{2}{3}\frac{p^{2}}{Q^{2}_{0}}\left[1+(\xi^{2}-1)\vz^{2} \right]}}{\sqrt{1+(\xi^{2}-1)\vz^{2}}} \left[ 1+ \sum_{m>0} \cos (m\phi) \right] \,\nonumber \\
&\times
\left[1+\sum_{n=1} \alpha_n \cos(n \cos^{-1}(\vz))\right]
\,.
\end{align}
$Q_{0}$ is the typical energy of hard gluons and we set $\xi=2$. 
Upper panels of \Fig{fig:attractor} show numerical solutions for the ratio of lowest moments in each sector, $L_{m+\delta_{s-}+2,m}/L_{m+\delta_{s-},m}$, for $(0,+)$,$(1,+)$,$(0,-)$ and $(2,+)$ sectors, with different initial conditions for $\alpha_n$ in different colors and truncating at $N=21$.
We observe that these solutions are independent of initial conditions before $\tau_R$ for $(0,+)$, $(1,+)$, and $(2,+)$ sectors, but not for $(0,-)$.
Our results for the $(0,+)$ sector are consistent with previous results in the literature, since $L_{20}/L_{00}$ is related to the scalar mode $p_L/\e$ which has been studied extensively~\cite{Romatschke:2017vte,Kurkela:2019set}.
Other sectors extend the observation of attractor behavior to vector and tensor components of the energy-momentum tensor, which are new in the literature. 
The evolution of the vector mode in the $(1,+)$ sector has both early- and late-time attractors, like the scalar mode. The vector mode in the $(0,-)$ sector, however, exhibits attractor behavior only when $\tau>\tau_R$. In the tensor mode $(2,+)$, attractor behavior is present throughout the whole evolution, but with the same power-law decay of initial conditions at early and late times. When attractor behavior is present, it is in good agreement with the evolution of the same ratio of moments for $\phi^G_{ms}$, shown in dashed black.

Features of the numerical solutions can be understood as consequences of the dominance of slow modes. Lower panels of \Fig{fig:attractor} show the energy gap between the ground state and the first excited state. Slow modes continuously dominate in the even sectors, resulting in attractor behavior at all times, while in the $(0,-)$ sector attractor behavior only occurs after the level crossing in Fig.~\ref{fig:eigenvalues}(c). In the even hydrodynamic modes $(0,+)$ and $(1,+)$, the time dependence of the energy gap differs at early and late times, resulting in power-law and exponential decays of initial conditions to the attractor. This also explains the observed power-law decay in the $(2,+)$ sector for all times.
Fig.~\eqref{fig:attractor} therefore
exhibits the diverse relations between hydrodynamic modes and slow modes in the far-from-equilibrium evolution that has already been foreshadowed in Fig.~\eqref{fig:eigenvalues}.

\begin{figure}
    \centering
    \includegraphics[width=\columnwidth]{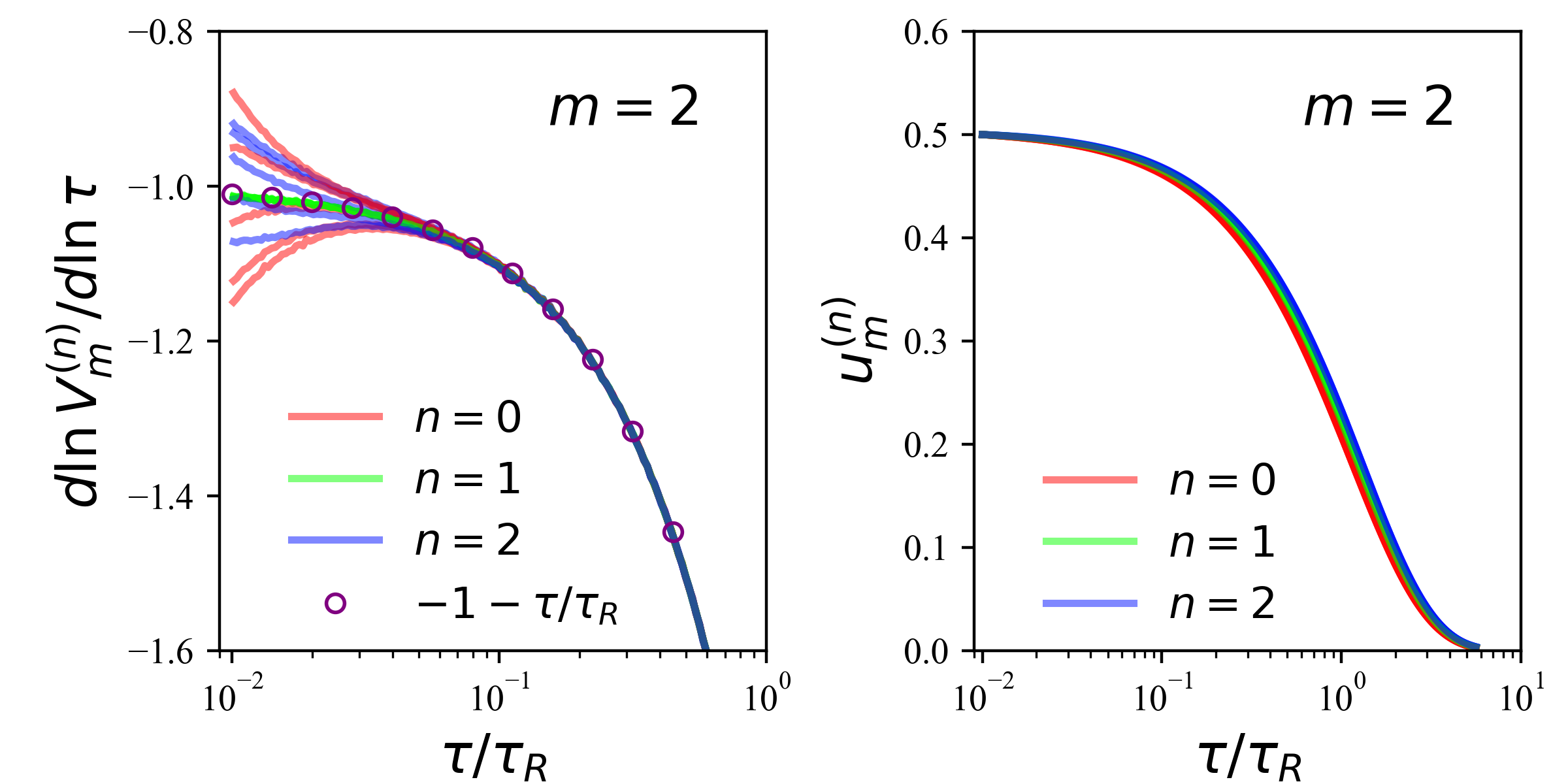}
    \caption{
    Evolution of moments related to the momentum-space anisotropy $u_2^{(n)}$. Left: rate of change of the generalized $V_2^{(n)}$ as a function of $\tau/\tau_R$ for numerical solutions (lines) and from Eqs.~\ref{eq:Vn} and~\ref{Em} (circles).
    Right: numerical solutions for the }evolution of $u_2^{(n)} = V_2^{(n)}/V_0^{(n)}$.
    \label{fig:Vn}
\end{figure}

\sect{Memory of initial momentum anisotropy}
In the very early stages of a heavy-ion collision, spatial gradients are much less important than the rapid longitudinal expansion, and \Eq{F-evo} thus provides a rough estimate of the evolution of initial momentum anisotropy. We quantify the momentum anisotropy from  $F(\hat p)$ through
$u^{(n)}_{m}\equiv V^{(n)}_{m}/V^{(0)}_{m}$, where $ V^{(n)}_{m}\equiv \int^{1}_{-1} d \vz\int \frac{d\phi}{4\pi}\, (\sqrt{1-\vz^{2}})^{n} \, F(\hp)\,\cos(m\phi)$.
For $n=2$, this is the standard definition of the momentum anisotropy.
For instance, $u^{(2)}_1\sim T^{0x}/\e$ is related to the rapidity-even dipole flow, while 
$u^{(2)}_{2}$ is
the momentum eccentricity ${\cal{E}}_{2p}\sim (T^{xx}-T^{yy})/\e$~\cite{Schenke:2019pmk}, 
which has been used 
as a proxy for the produced hadron elliptic flow~\cite{Kurkela:2019kip}. For higher $n$, $u_m^{(n)}$ is still related to anisotropy but with different momentum weights.

The $V^{(n)}_{m}$ are modes in the $(m,+)$ sector.
The evolution of $V^{(n)}_{m}$ is therefore determined by the ground state as
\begin{equation}
\label{eq:Vn}
d \ln V_m^{(n)}/d\ln \tau\sim -\tau E_{m+}^G\,,
\end{equation}
and we anticipate that arbitrary initial conditions attract to this evolution. 
This is verified in Fig.~\ref{fig:Vn}, where
the rate of change of $V^{(n)}_{2}$ quickly approaches $-\tau E^{G}_{2+}$ (given by Eq.~\ref{Em}), for different initial conditions and independent of the weight $n$. 
This implies an evolution equation for the initial momentum eccentricity,
\begin{align}
\label{um-evo}
\partial_\tau \ln {\cal E}_{2p}\sim
    \pd_{\tau}\ln u^{(2)}_{2}=-\left(E^{G}_{2+}-E^{G}_{0+}\right)\,. 
\end{align}
\Eq{um-evo} is a non-trivial consequence of the attractor behavior of $V_m^{(n)}$. Furthermore,
\Eq{um-evo} has important implications for the evolution of initial momentum anisotropy in a far-from-equilibrium plasma. In the free-streaming case, $E_{m+}^G$ are degenerate for all $m$ so the anisotropy in Eq.~\ref{um-evo} does not evolve. However, there is a small difference between the ground states of $(m,+)$ and $(0,+)$ when $m \geq 2$, namely $E^{G}_{m+}-E^{G}_{0+}\sim 1/\tau_R$
(see Eq.~\eqref{Em}). 
Consequently, initial momentum anisotropies
should survive up to the relaxation time $\tau_{R}$, which is a main result of this letter. 
The decay of initial momentum eccentricity is explicitly illustrated in Fig.~\eqref{fig:Vn} as an example. 
The
dominance of slow modes in the even sectors additionally implies that Eq.~\eqref{um-evo} applies to momentum anisotropies of arbitrary order $m \geq 2$. 
The evolution of the triangularity $m=3$ is similar (see Supplemental material), confirming the applicability of our analysis to higher moments of the distribution function.

In heavy-ion collisions, momentum anisotropy is also generated from the initial geometry on the time scale $\tau_{\cal E}\sim R/c_s$~\cite{Ollitrault:2007du,Kurkela:2019kip} that is determined by the characteristic transverse size of the medium $R$ and the sound velocity $c_s$.
Therefore the memory of the initial anisotropy described here could become important by comparison when $\tau_{\cal E}/\tau_R<1$.
Estimating the relaxation time from $\tau_R=4\pi(\eta/s)/T$ and using the relation between initial entropy and charged particle multiplicity per rapidity $y$, $dS/dy\sim \nu T^3 \tau_0 \pi R^2\sim 7.14\, dN_{\rm ch}/dy$~\cite{Churchill:2020uvk}, gives   
\beq
\frac{\tau_{\mathcal{E}}}{\tau_R}
\sim \left(\frac{R}{\tau_0}\right)^{1/3} \left(\frac{7.14}{\nu \pi} \frac{dN_{\rm ch}}{dy}\right)^{1/3} \frac{1}{4\pi(\eta/s)\,c_{s}}\, .
\eeq
With a ballpark value $\nu=40$ for the number of effective degrees of freedom and $\eta/s=0.2$, we conclude that in small colliding systems with $R/\tau_0\sim O(1)$, the initial momentum anisotropy can be important for the observed anisotropic flow when $dN_{\rm ch}/dy\lesssim 20$.

\sect{Discussion}%
We have studied slow modes in a longitudinally-expanding plasma using kinetic theory in the relaxation-time approximation.
While late-time slow modes are unambiguously hydrodynamic modes, the relation between early-time slow modes and hydrodynamic ones is rather diverse. At early times when the system is very far from equilibrium,
not all hydrodynamic modes are slow modes and not all slow modes are hydrodynamic modes.
Importantly, we found early-time slow modes associated with the momentum anisotropy that dominate the evolution of the anisotropy and are not related to hydrodynamic modes. 
The rate of change of those early-time non-hydrodynamic slow modes and (parity even) hydrodynamic modes is degenerate in the early-time limit. The novel result of these two features is
that momentum-space anisotropy from the initial state survives up to relaxation time $\tau_{R}$.

These features are driven not by collisions but by rapid longitudinal expansion, so we anticipate that these conclusions are general for systems where free-streaming expansion dominates at early times. In the future it will be interesting to use this type of approach to study the early-time phase of strongly-coupled systems.
This study
further 
motivates future work to study the interplay between momentum anisotropy from the initial state and generated by spatial gradients.

There are a number of aspects of our results that are of broader interest. Demonstrating the emergence of non-hydrodynamic slow modes in rapidly-expanding plasmas may provide insight into finding slow modes in other systems.
The non-hydrodynamic slow modes discussed here describe the shape of phase space distribution. 
Interestingly, 
the shape of the Fermi surface is identified as a slow mode in the effective theory for Fermi liquids~\cite{Delacretaz:2022ocm} and has recently been proposed as a low-energy degree of freedom in the extended hydrodynamics for fractional quantum Hall states~\cite{Son:2019qlm}.

\sect{Acknowledgements}
We are grateful for valuable discussions with Aleksas Mazeliauskas, Krishna Rajagopal, S\"oren Schlichting, Michael Strickland, and Urs Wiedemann.
WK is supported by the US Department of Energy through the Office of Nuclear Physics and the LDRD program at Los Alamos National Laboratory. Los Alamos National Laboratory is operated by Triad National Security, LLC, for the National Nuclear Security Administration of U.S. Department of Energy (Contract No. 89233218CNA000001). 
LY is partly supported by National Natural Science Foundation of China through grant No. 11975079.
YY acknowledges the support from the Strategic Priority Research Program of Chinese Academy of Sciences, Grant No. XDB34000000 and NSFC under grant No.12175282.

\bibliographystyle{apsrev}
\bibliography{ref}

\newpage

\appendix

\onecolumngrid

\section*{Supplemental Material}

\subsection{Expressions for $H_{ms}$ and its properties}

Here we elaborate on the technical steps leading to moment equation~\eqref{psims}. 
We begin with an alternative expression for Eq.~\eqref{F-evo} 
\begin{align}
\label{F-evo0}
    \pd_{\tau}\, F= - \frac{1}{\tau}\le[4\vz^{2}-\vz\, (1-\vz^{2})\,\pd_{\vz}\ri]\, F- \,C[F]\, .
\end{align}
To determine the evolution equation for the moments $L_{lm}$, 
we multiply $X_{lm}\equiv(-1)^{m}\, P^{m}_{l}(\vz)\,\cos(m\phi)$ on both sides of Eq.~\eqref{F-evo} and integrate over the solid angle. 
Using the definition of the moments~\eqref{moments}, 
we find
\begin{align}
\label{L-evo}
    \pd_{\tau}L_{lm}(\tau)=-\frac{1}{\tau}\langle X_{lm}(\hp) \le[4\vz^{2}-\vz\, (1-\vz^{2})\,\pd_{\vz}\ri]\,F(\hp)\rangle -
    \langle X_{lm}(\hp) C[F(\hp)]\rangle\, . 
\end{align}
Here and throughout we denote the average over the solid angle by
\begin{align}
    \langle\ldots \rangle \equiv \frac{1}{2}\int^{1}_{-1} d\vz\, \int^{2\pi}_{0}\frac{d\phi}{2\pi} (\ldots)\,. 
\end{align}
Next, we express $F$ in terms of moments $L_{lm}$
\begin{align}
\label{F-exp}
    F(\tau,\hp)= \sum_{l m}\, \frac{L_{lm}(\tau)}{N_{lm}}\, X_{lm}(\hp)\, , 
\end{align}
where we have used orthonormality relations
\begin{align}
\label{orth}
&\,    \langle X_{lm}(\hp)X_{l'm'} \rangle= N_{lm}\delta_{ll'}\delta_{mm'}\, , 
\qquad
N_{lm}= \frac{1}{2(2l+1)}\frac{(l+m)!}{(l-m)!}\left(1+\delta_{m0}\right)\, . 
\end{align}
To evaluate the first term on the right-hand side of \eqref{L-evo}, we use the recursion relations of Legendre polynomials
\begin{align}
\label{P-recur}
xP_{l}^m(x) &= a^{m}_{l}\, P_{l+1}^m(x) + b^{m}_{l}P_{l-1}^m(x),\, 
\qquad
(1-x^2) \frac{d}{dx} P_{l}^m(x) = -c^{m}_{l}\,P_{l+1}^m(x) + d^{m}_{l}\,P_{l-1}^m(x)
\end{align}
where the coefficients are non-zero for $l\geq 0$
\begin{eqnarray}
\label{ab-def}
a_{l}^m &=& \frac{l-m+1}{2l+1}\, ,\qquad
b^{m}_{l}=\frac{l+m}{2l+1},\, 
\qquad
c^{m}_{l}=\frac{l(l-m+1)}{2l+1}\, ,
\qquad
d^{m}_{l}=\frac{(l+1)(l+m)}{2l+1}\, . 
\end{eqnarray}
We consequently have
\begin{align}
\label{unintTerm1}
\left[-4\vz^2+\vz(1-\vz^{2})\pd_{\vz}\right]X_{l,m}
=&-\left(4 a^{m}_{l}+c^{m}_{l}\right)a^{m}_{l+1}\, X_{l+2,m}
-\left[\left(4a^{m}_{l}+c^{m}_{l}\right)b^{m}_{l+1}+\left(4b^{m}_{l}-d^{m}_{l}\right)a^{m}_{l-1} \right] X_{l,m}\nonumber\\& -\left(4b^{m}_{l}-d^{m}_{l}\right)b^{m}_{l-1}\,X_{l-2,m}
\end{align}
The above expression combined with Eq.~\eqref{F-exp} gives
\begin{align}
\label{Term1}
\langle X_{lm} \left[-4\vz^2+\vz(1-\vz^{2})\pd_{\vz}\right]F(\hp)\rangle    
=& 
-\frac{N_{l,m}}{N_{l+2, m}}\left(4b^{m}_{l+2}-d^{m}_{l+2}\right)b^{m}_{l+1}\,L_{l+2,m}
-\left[\left(4 a^{m}_{l}+c^{m}_{l}\right)b^{m}_{l+1}+\left(4b^{m}_{l}-d^{m}_{l}\right){a_{l-1}^m}\right]\,L_{l,m}
\nonumber \\
&
-
\frac{N_{l,m}}{N_{l-2, m}}\left(4 a^{m}_{l-2}+c^{m}_{l-2}\right)\,a^{m}_{l-1}\, L_{l-2, m}\, . 
\end{align}
We now turn to the second term on the right-hand side of Eq.~\eqref{L-evo}. Since the collision integral preserves energy and momentum, we must have 
\begin{align}
    \langle X_{lm}(\hp) C[F(\hp)]\rangle=0\, , 
    \qquad (lm)=(lm)_{H}
\end{align}
where $(lm)_{H}$ means $(lm)$ belongs to one of $(00),(10),(11)$ associated with hydrodynamic moments. For the collision integral~\eqref{RTA}, moments with different $lm$ do not mix and we simply have
\begin{align}
\label{Term2}
    \langle X_{lm}(\hp) C[F(\hp)]\rangle=\frac{1}{\tau_{R}}\, \left(1-\delta_{lm,(lm)_{H}}\right)\, . 
\end{align}
Explicit expressions for the evolution of the moments in Eq.~\eqref{L-evo} can be read from Eqs.~\eqref{Term1} and ~\eqref{Term2}.
As discussed in the main text, since moments with different $m$ and even or odd values of $l+m$ do not mix under Eqs.~\eqref{Term1} and ~\eqref{Term2}, we reformulate Eq.~\eqref{L-evo} as a matrix equation for the evolution of a vector of moments $\psi_{ms}$, indexed by $m$ and $s = +(-)$ for even(odd) $l+m$.
For example, if we only keep the first $5$ components in $\psi_{ms}$, the corresponding matrices are
\begin{align}
H_{0+}&=\frac{1}{\tau}\left(
\begin{array}{ccccc}
 \frac{4}{3} & \frac{2}{3} & 0 & 0 & 0 \\
 \frac{8}{15} & \lambda +\frac{38}{21} & -\frac{12}{35} & 0 & 0 \\
 0 & \frac{8}{7} & \lambda +\frac{136}{77} & -\frac{10}{11} & 0 \\
 0 & 0 & \frac{240}{143} & \lambda +\frac{58}{33} & -\frac{56}{39} \\
 0 & 0 & 0 & \frac{112}{51} & \lambda +\frac{100}{57} \\
\end{array}
\right)
\qquad
H_{1+}=\frac{1}{\tau}\left(
\begin{array}{ccccc}
 1 & 0 & 0 & 0 & 0 \\
 \frac{12}{7} & \lambda +\frac{5}{3} & -\frac{8}{21} & 0 & 0 \\
 0 & \frac{70}{33} & \lambda +\frac{67}{39} & -\frac{120}{143} & 0 \\
 0 & 0 & \frac{168}{65} & \lambda +\frac{383}{221} & -\frac{112}{85} \\
 0 & 0 & 0 & \frac{990}{323} & \lambda +\frac{207}{119} \\
\end{array}
\right)
\\
H_{0-}&= \frac{1}{\tau} \left(
\begin{array}{ccccc}
 2 & 0 & 0 & 0 & 0 \\
 \frac{6}{7} & \lambda +\frac{16}{9} & -\frac{40}{63} & 0 & 0 \\
 0 & \frac{140}{99} & \lambda +\frac{206}{117} & -\frac{168}{143} & 0 \\
 0 & 0 & \frac{126}{65} & \lambda +\frac{388}{221} & -\frac{144}{85} \\
 0 & 0 & 0 & \frac{792}{323} & \lambda +\frac{626}{357} \\
\end{array}
\right)
\,, 
\qquad
H_{2+}=\frac{1}{\tau}
\left(
\begin{array}{ccccc}
 \lambda +\frac{6}{7} & -\frac{2}{35} & 0 & 0 & 0 \\
 \frac{20}{7} & \lambda +\frac{116}{77} & -\frac{4}{11} & 0 & 0 \\
 0 & \frac{448}{143} & \lambda +\frac{18}{11} & -\frac{10}{13} & 0 \\
 0 & 0 & \frac{60}{17} & \lambda +\frac{32}{19} & -\frac{392}{323} \\
 0 & 0 & 0 & \frac{528}{133} & \lambda +\frac{746}{437} \nonumber\\
\end{array}
\right)
\end{align}
where $\lambda=\tau/\tau_{R}$.

\subsection{The ground state in the free-streaming limit}

In this section, we discuss the ground state of $H_{m+}$ in the free-streaming limit. 
In this limit, we can ignore the collision integral in the kinetic equation and consider
\begin{align}
\label{f}
    \pd_{\tau}f-\frac{p_{z}}{\tau}\pd_{p_{z}}f=0\, . 
\end{align}
For an arbitrary initial condition $f(\tau_{I},p_{z},p_{\perp})=f_{I}(p_{z},p_{\perp})$,
the solution to \eqref{f} can be found analytically
\begin{align}
    f(\tau,p_{z},p_{\perp})=f_{I}\left(\frac{\tau}{\tau_{I}}p_{z},p_{\perp}\right)\, . 
\end{align}
When $\tau\gg \tau_{I}$ but still smaller than $\tau_{R}$ (which can be satisfied when $\tau_{I}\to 0$), the distribution should become sharply peaked at $\vz=0$, assuming $f_{I}$ has finite support in momentum space. 
On the other hand, the moments should be described by the ground state for sufficiently large $\tau$ . 
This means that free-streaming ground state takes the form
\begin{align}
\label{phi-F}
    \phi^{G}_{m+}= \left( P^{m}_{m}(0), P^{m}_{m+2}(0),\ldots \right)\, . 
\end{align}

We will now verify that Eq.~\eqref{phi-F} is indeed an eigenstate with eigenvalue $1/\tau$, independent of $m$. 
According to Eq.~\eqref{unintTerm1}, to prove this we need to show
\begin{align}
\label{relation}
&\,    \frac{N_{l,m}}{N_{l+2, m}}\left(4b^{m}_{l+2}-d^{m}_{l+2}\right)b^{m}_{l+1}\,P^{m}_{l+2}(0)+
\left[\left(4 a^{m}_{l}+c^{m}_{l}\right)b^{m}_{l+1}+\left(4b^{m}_{l}-d^{m}_{l}\right){a_{l-1}^m}\right]\,P^{m}_{l}(0)
\nonumber 
\\
&+
\frac{N_{l,m}}{N_{l-2, m}}\left(4 a^{m}_{l-2}+c^{m}_{l-2}\right)\,a^{m}_{l-1}\, P^{m}_{l-2}(0)= P^{m}_{l}(0)\,,
\end{align}
or
\begin{equation}
\label{relation-0}
\frac{N_{l,m}}{N_{l+2, m}}\left(4b^{m}_{l+2}-d^{m}_{l+2}\right)b^{m}_{l+1}\,\frac{P^{m}_{l+2}(0)}{P^{m}_{l}(0)}+
\left[\left(4 a^{m}_{l}+c^{m}_{l}\right)b^{m}_{l+1}+\left(4b^{m}_{l}-d^{m}_{l}\right){a_{l-1}^m}\right]+
\frac{N_{l,m}}{N_{l-2, m}}\left(4 a^{m}_{l-2}+c^{m}_{l-2}\right)\,a^{m}_{l-1}\, \frac{P^{m}_{l-2}(0)}{P^{m}_{l}(0)}= 1\, . 
\end{equation}
From the recursive relation \eqref{P-recur}, we have
\begin{align}
\label{P-0}
    P^{m}_{l+2}(0)= -\frac{b_{l+1}}{a_{l+1}}\, P^{m}_{l}(0)\, , 
    \qquad
    P^{m}_{l-2}(0)= -\frac{a_{l-1}}{b_{l-1}}\, P^{m}_{l}(0)\, .
\end{align}
Substituting \eqref{P-recur} into \eqref{relation-0} and using the definitions~\eqref{ab-def}, we can verify algebraically that Eq.~\eqref{relation-0} is satisfied when $l+m$ is even. 
We note that the eigenvalue is independent of $m$, meaning there is an associated slow mode for each $m$.

\subsection{The derivation of Eq.~\eqref{Em}}
Finally, for the purpose of deriving Eq.~\eqref{Em}, we will consider the contributions from Eqs.~\eqref{Term1} and \eqref{Term2} separately by writing $H_{ms}=\frac{1}{\tau}(\sH_{F,ms}+\sH_{C,ms})$. For $m \geq 2$ and $s=+$,
we observe from Eq.~\eqref{Term2} that $\sH_{C,m+}$ is a unit matrix and hence commutes with $\sH_{F,m+}$.
Therefore if $\phi^{G}_{m+}$ is an eigenmode of $\sH_{F,m+}$ with eigenvalue $1$, it is also an eigenmode of $\sH_{C,m+}$ with eigenvalue $\lambda$. 
This gives
\begin{align}
    H_{m+}\, \phi^{G}_{m+}= \frac{1}{\tau}\le[ \sH_{F,m+}+\lambda\ri] \, \phi^{G}_{m+}
    =\left( \frac{1}{\tau}+\frac{1}{\tau_{R}}\right)\, \phi^{G}_{m+},
\end{align}
which verifies Eq.~\eqref{Em}.

\subsection{Evolution of momentum-space triangularity}

\begin{figure}
    \centering
    \includegraphics[width=.5\textwidth]{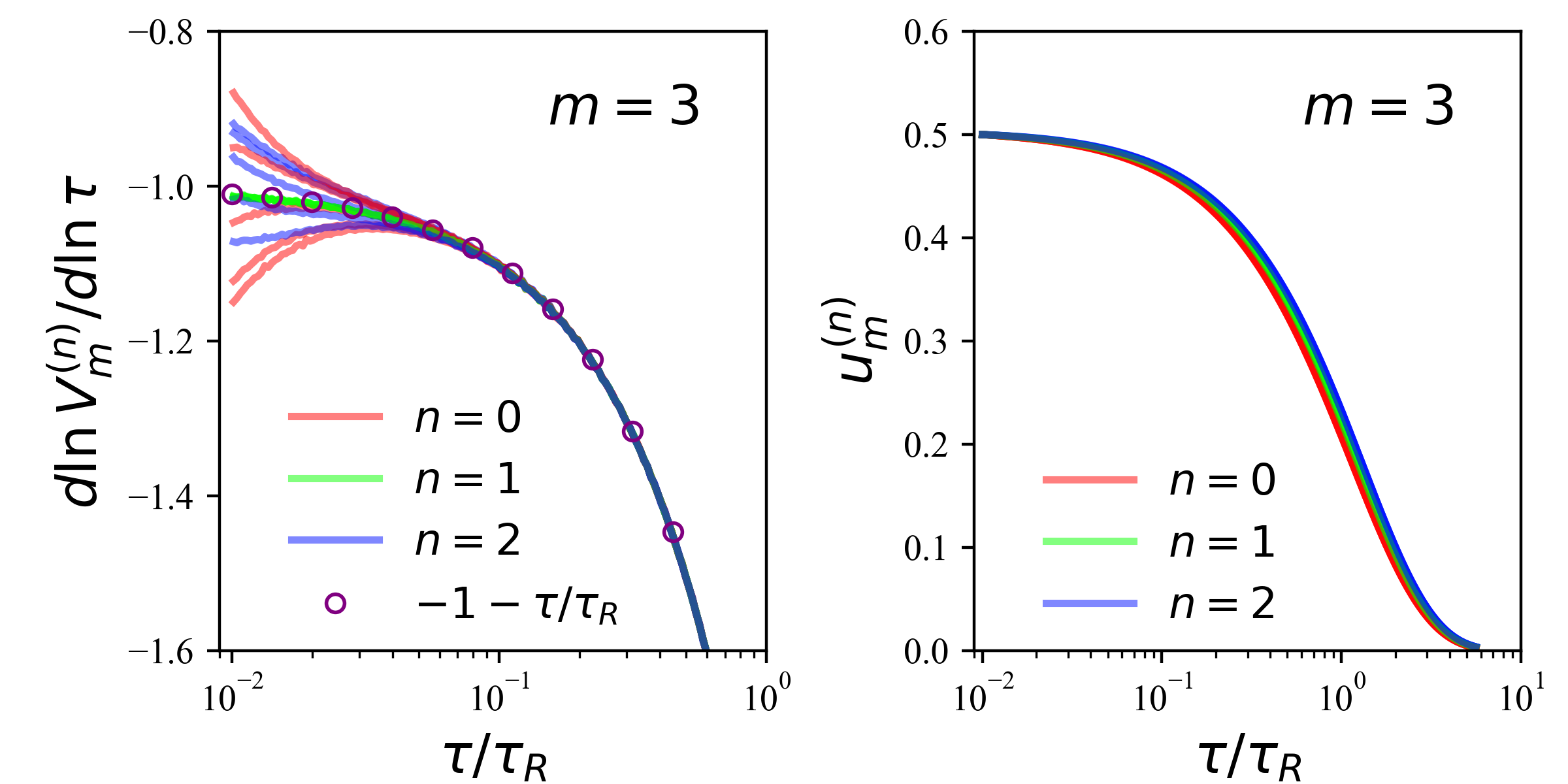}
    \caption{
    Left panel: rate of change of the generalized $V_3^{(n)}$ as a function of $\tau/\tau_R$ for numerical solutions (lines) and from Eq.~\ref{Em} (circles).
    Right panel: evolution of $u_3^{(n)} = V_3^{(n)}/V_0^{(n)}$.
    }
    \label{fig:V3}
\end{figure}

In Fig.~3 of the main text, we exhibit the evolution of the momentum-space ellipticity, $u_2$. To complement these results, 
in Fig.~\ref{fig:V3} we show the evolution of the momentum-space triangularity $u_3$. 
We observe similar behavior to $u_2$. 
This confirms our anticipation that the dominance of early-time slow modes also applies to higher harmonics of the momentum distribution.

\end{document}